\documentclass[a4paper]{jpconf}
\usepackage{graphicx}
\begin{document}
\title{Test beam studies of the TRD prototype filled with different gas mixtures based on Xe, Kr, and Ar}
\author{
 E~Celebi$^{4}$\footnote[8]{To whom any correspondence should be addressed. },
 T~Brooks$^{3}$,
 M~Joos$^{3}$,
 C~Rembser$^{3}$,
 S~Gurbuz$^{4}$,
 S~A~Cetin$^{5}$,
 S~P~Konovalov$^{2}$,
 V~O~Tikhomirov$^{1,2,}$,
 K~Zhukov$^{2}$,
 K~A~Fillipov$^{1,2}$,
 A~Romaniouk$^{1}$,
 S~Yu~Smirnov$^{1}$,
 P~E~Teterin$^{1}$,
 K~A~Vorobev$^{1}$,
 A~S~Boldyrev$^{6}$,
 A~Maevsky$^{6}$,
 and D~Derendarz$^{7}$}

\address{$^1$ National Research Nuclear University MEPhI (Moscow Engineering
Physics Institute), Kashirskoe highway 31, Moscow, 115409, Russia}
\address{$^2$ P. N. ~Lebedev Physical Institute of the Russian Academy of
Sciences, Leninsky prospect 53, Moscow, 119991, Russia}
\address{$^3$ CERN, the European Organization for Nuclear Research, CH-1211 Geneva 23, Switzerland}
\address{$^4$ Bo\u{g}azi\c{c}i University,34342 Bebek/Istanbul Turkey}
\address{$^5$ Istanbul Bilgi University, High Energy Physics Research Center, Eyup, Istanbul, 34060, Turkey}
\address{$^{6}$ Skobeltsyn Institute of Nuclear Physics Lomonosov Moscow State
 University, Moscow, Russia}
\address{$^{7}$ Institute of Nuclear Physics Polish Academy of Sciences, Krakow, Poland}

\ead{Emre.Celebi@cern.ch}

\begin{abstract}
Towards the end of LHC Run1, gas leaks were observed in some parts of the Transition Radiation Tracker (TRT) of ATLAS. Due to these leaks, primary Xenon based gas mixture was replaced with Argon based mixture in various parts. Test-beam studies with a dedicated Transition Radiation Detector (TRD) prototype were carried out in 2015 in order to understand transition radiation performance with mixtures based on Argon and Krypton. We present and discuss the results of these test-beam studies with different active gas compositions. 
\end{abstract}

\section{Introduction}

 TRT is the outermost part of the ATLAS Inner Detector. It is a gaseous tracker system consisting of ~300 000 small diameter(4 mm) drift tubes called straws. Along with its tracking capability TRT also helps Particle Identification (PID) by measuring transition radiation. TRT compares signals from the straws with two different thresholds; one being low the other being high. When the deposited energy results in a signal which is over the high level treashold in the straw, it is considered as a HL hit. HL hits indicate the transition radiation.

\begin{figure}[h]
\begin{minipage}{14pc}
\includegraphics[width=14pc]{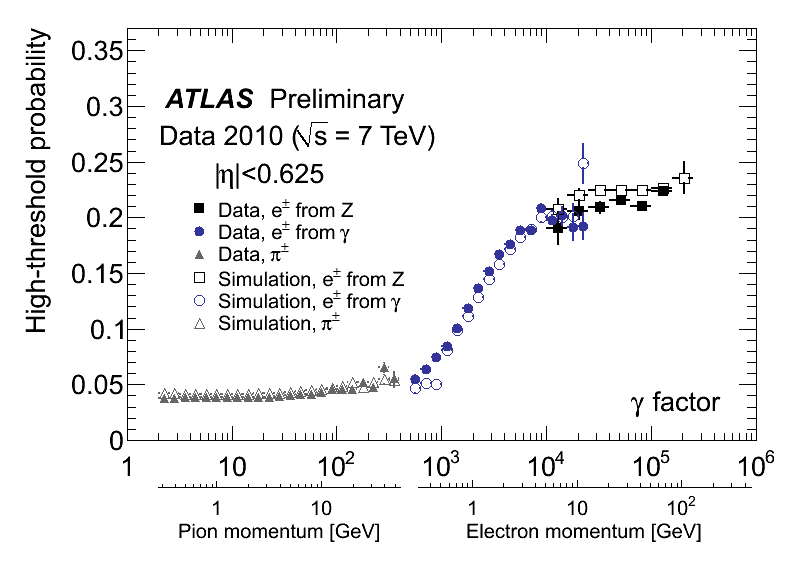}
\caption{\label{fig:Turnon}HL threashold probability versus the Lorentz factor in the barrel region of the TRT~\cite{ATLAS-CONF-2011-128}. }
\end{minipage}\hspace{2pc}%
\begin{minipage}{14pc}
\includegraphics[width=14pc]{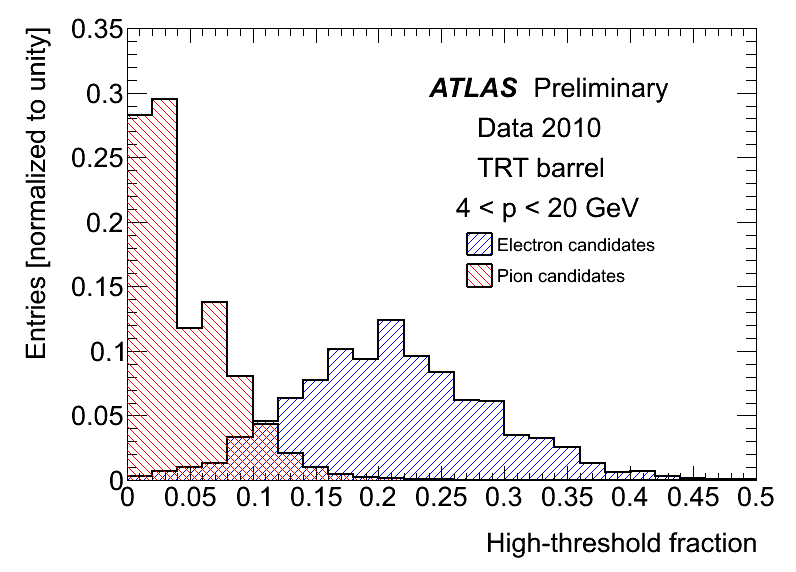}
\caption{\label{fig:HL_LL}HL hit fraction for electron and pion candidates in the barrel region of the TRT\cite{ATLAS-CONF-2011-128}. }
\end{minipage}
\end{figure}

 Transition radiation occurs when a highly relativistic charged particle crosses the boundary between media with different dielectric properties. The radiated energy with transition radiation(TR) is roughly proportional to Lorentz factor ($\gamma$). The spectrum hardens as the particle velocity increases \cite{PDG-2014}.  As a result HL hit probability changes with the Lorentz factor(Figure~\ref{fig:Turnon}). Since their $\gamma$ is different, HL fraction on the track for electrons and pions with the same momentum differs as shown in Figure~\ref{fig:HL_LL}. Therefore PID depends on threshold setting and the detecting medium.

Due to some unforeseeable effects cracks developed in the TRT gas pipes and some of the damage was impossible to repair. Hence in Run 2, Argon based gas mixture was used in critically effected parts instead of the Xenon based gas mixture. This affected the PID performance since Argon TR capture efficiency is significantly lower.

 In this study, we examine the TR detection performance the transition radiation detector (TRD) prototype filled with different gas mixtures based on Xenon , Krypton, and Argon.The results of some other studies performed with the TRD prototype can be also found in \cite{vlad}, \cite{graphen} presented at this Conference.

\section{Motivation Of The Study With Krypton Gas Mixture}

\begin{figure}[h]
\includegraphics[width=14pc]{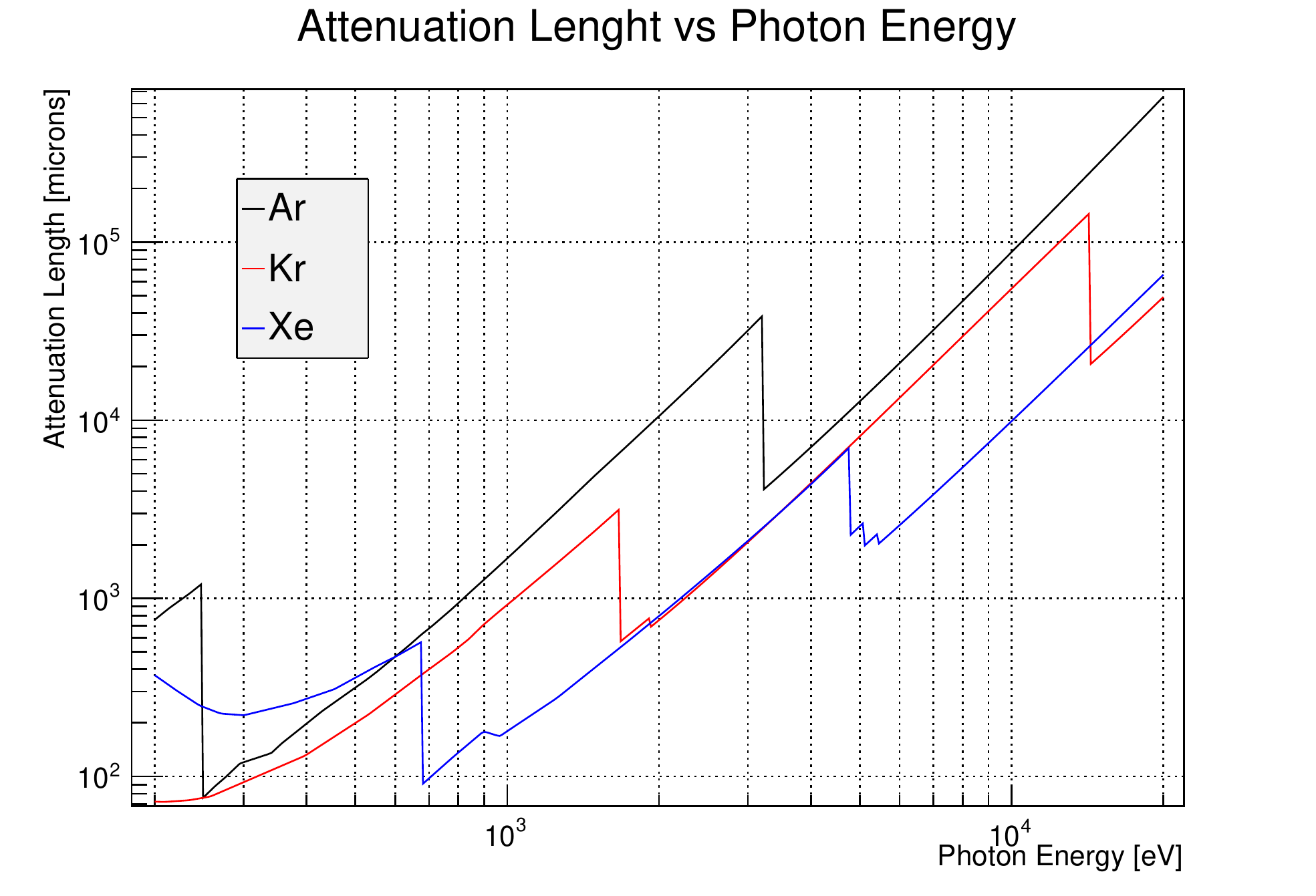}\hspace{2pc}%
\begin{minipage}[b]{14pc}\caption{\label{AttenL} X-ray attenuation lenght, data taken from CXRO web page. ~\cite{CXRO}}
\end{minipage}
\end{figure}

 The TR spectrum is roughly from 4 keV to 20 keV and in this energy range 
the Xenon absorption crossection is higher than Argon and Krypton for photons upto ~14keV. 
After 14keV Krypton K-shell contribution to crossection decreases the absorption lenght value for Krypton to lower level than the rest (Figure\ref{AttenL}). 
When a 14 keV photon iteracts with Krypton atom, there is 65\% probability that a 12 keV escape photon is emmited. Yet there is roughly 2 keV energy left in the straw and combined with the energy deposited by the particle itself (dE/dx) which all together contributes to the energy range where TR expected. 

As a result the Krypton mixture(after threshold optimization) seems to be a good candidate for PID purposes.

\section{2015 TRT Test Beam Setup}

\begin{figure}[h]
\includegraphics[width=25pc]{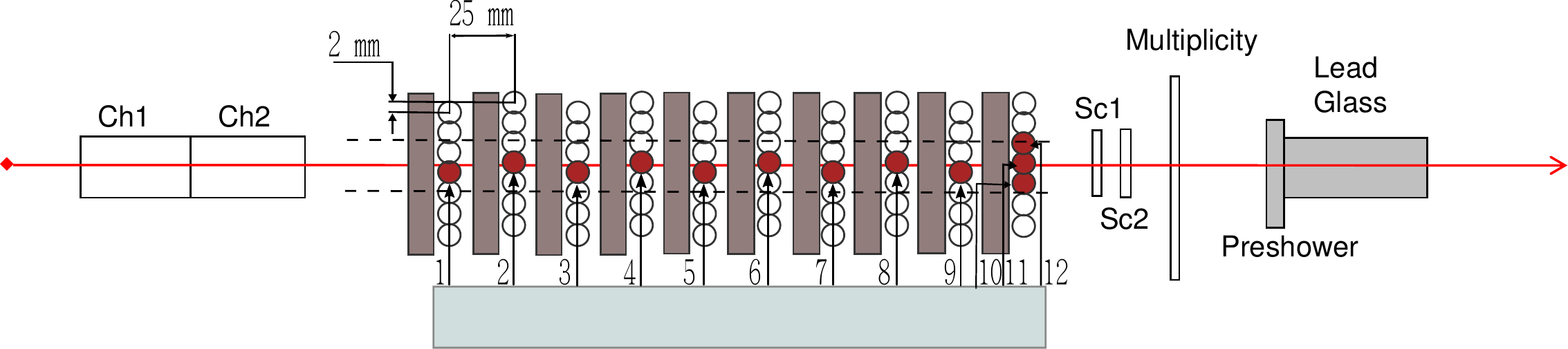}\hspace{2pc}%
\begin{minipage}[b]{10pc}\caption{\label{setup}Schema of the TRT Test Beam setup.}
\end{minipage}
\end{figure}

 The 2015 TRT test beam was held at CERN (SPS north area H8) during 26$^{th}$ May - 1 $^{st}$ June. 
 The main aim of the effort was to asses the TR photon capture performance of the Krypton based gas mixture. 
 A transition radiation detector (TRD) made out of TRT straws was used with radiators similar to radiator materials used in TRT. 
 In Figure \ref{setup}, circles represent the straws ande the rectangles represent the radiators infront of the straws. 
 The two scintillators each about the size of the beam profile (Sc1,Sc2) were used for triggering; their coincidence was the trigger condition.

 Electron-pion mixed beams of 20 GeV were used during the test beam study. There were two beam configurations; one being the electron-rich configuration and the other being pion-rich configuration. 
 Measurement with Xenon, Argon and Krypton based gas mixtures were performed with roughly 70\% noble gas, 27\% CO$_2$, 3\% O$_2$ composition. 
 Standart Fe55 calibration was done for each straws. 
 Gas gain controlled with an accuracy of about 2\%. 
 Different radiators and arrangements such as fibre and foil (Polypropylene) radiators as well as radiator-absent case were set up during data taking.

\section{Particle Identification}
\begin{figure}[h]
\begin{minipage}{16pc}
\includegraphics[width=16pc]{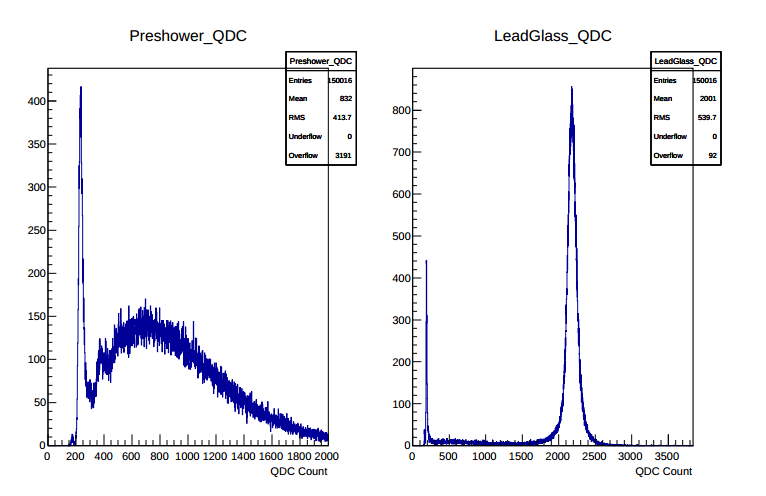}
\caption{\label{fig:erich} Preshower and Lead-glass QDC values for electron-rich beam. }
\end{minipage}\hspace{2pc}%
\begin{minipage}{16pc}
\includegraphics[width=16pc]{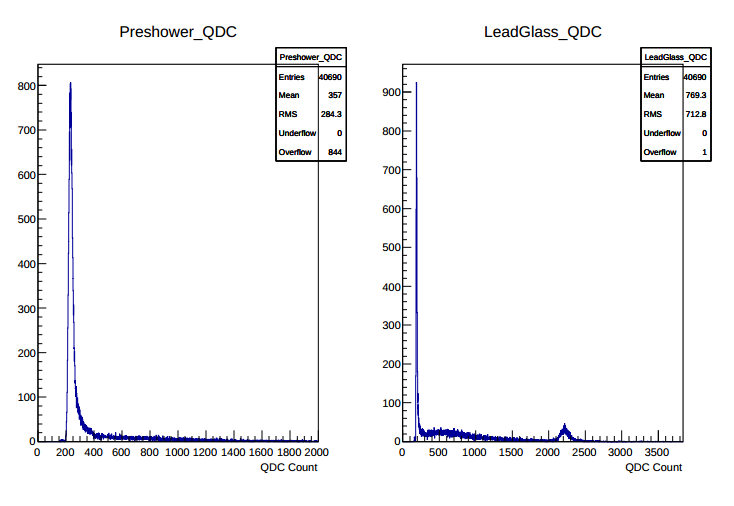}
\caption{\label{fig:pirich}Preshower and Lead-glass QDC values for pion-rich beam. }
\end{minipage}
\end{figure}

 Most of the data taking was with electron-rich beams however we had few pion-rich runs as well. The lead-glass and preshower detectors were used to identify pions and electrons. 
 The lack of different-composition runs beyond the two types mentioned above overrules the possibility of a quantitative study of the beam composition in an attempt to determine the amount of possible contamination from other particles, such as protons and kaons. 
 If these contaminations were of a significant level, we would expect to see the tail of a decaying function underneath the sharp electron peaks in the leadglass calorimeter QDC count histograms. (Figures \ref{fig:erich}\&\ref{fig:pirich})
 
 The identification selection criteria are as follows;
\begin{itemize}

\item Particles with the leadglass QDC values between 2000-2500 and preshower QDC values between 400-3500 are considered as electrons. 
\item Particles with the leadglass QDC values between 0-1500 and preshower QDC values between 200-300 are considered as pions. 
\end{itemize}

Particle selection purity with this selection is around 10$^{-3}$. 

\section{Differential and Integral Spectrums for Xe, Ar, Kr}

\begin{figure}[h]
\begin{minipage}{15pc}
\includegraphics[width=15pc]{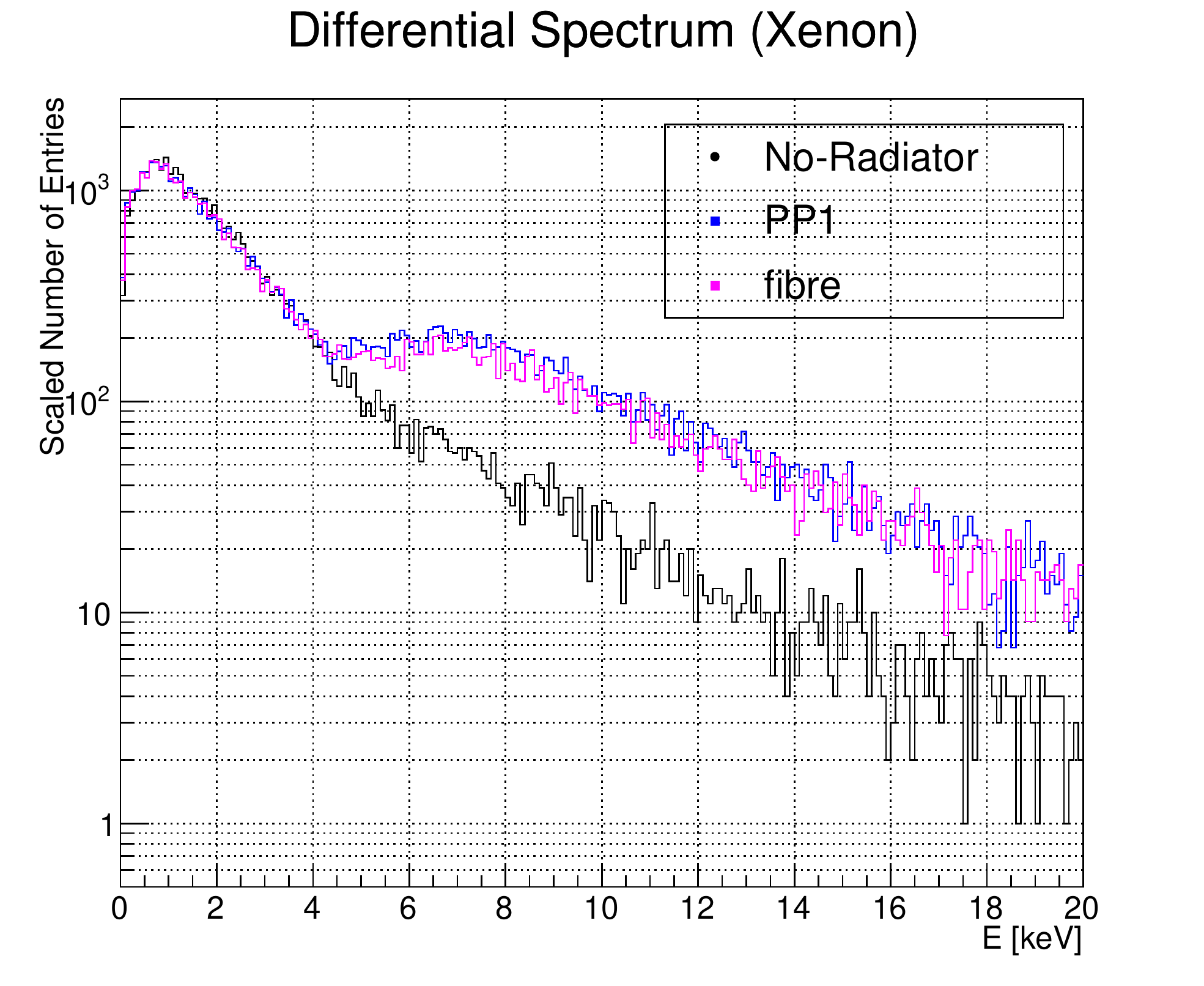}
\caption{\label{diff_xe}Xenon differential spectrum. }
\end{minipage}\hspace{2pc}%
\begin{minipage}{15pc}
\includegraphics[width=15pc]{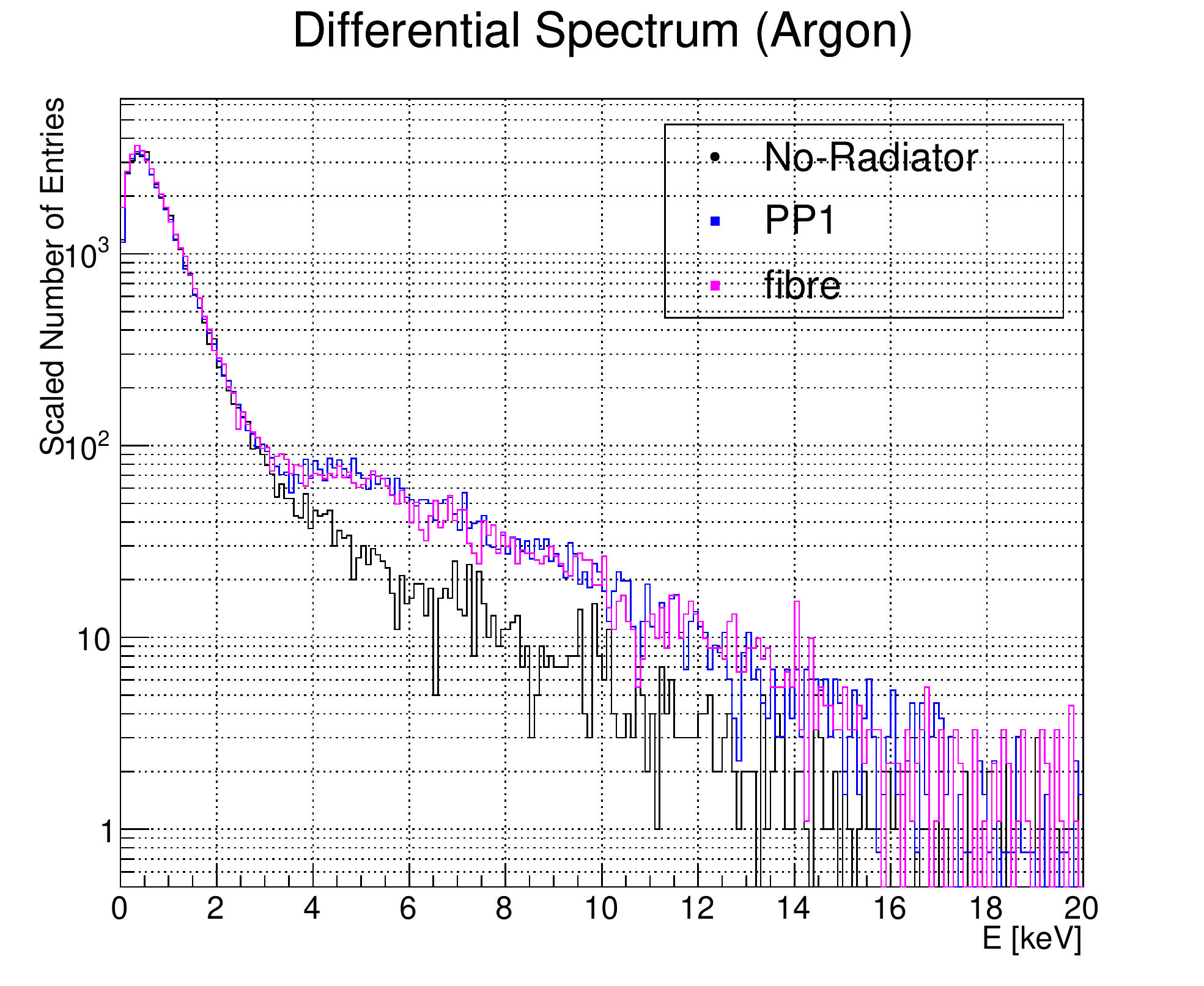}
\caption{\label{diff_ar}Argon differential spectrum. }
\end{minipage}
\end{figure}

\begin{figure}[h]
\begin{minipage}{15pc}
\includegraphics[width=15pc]{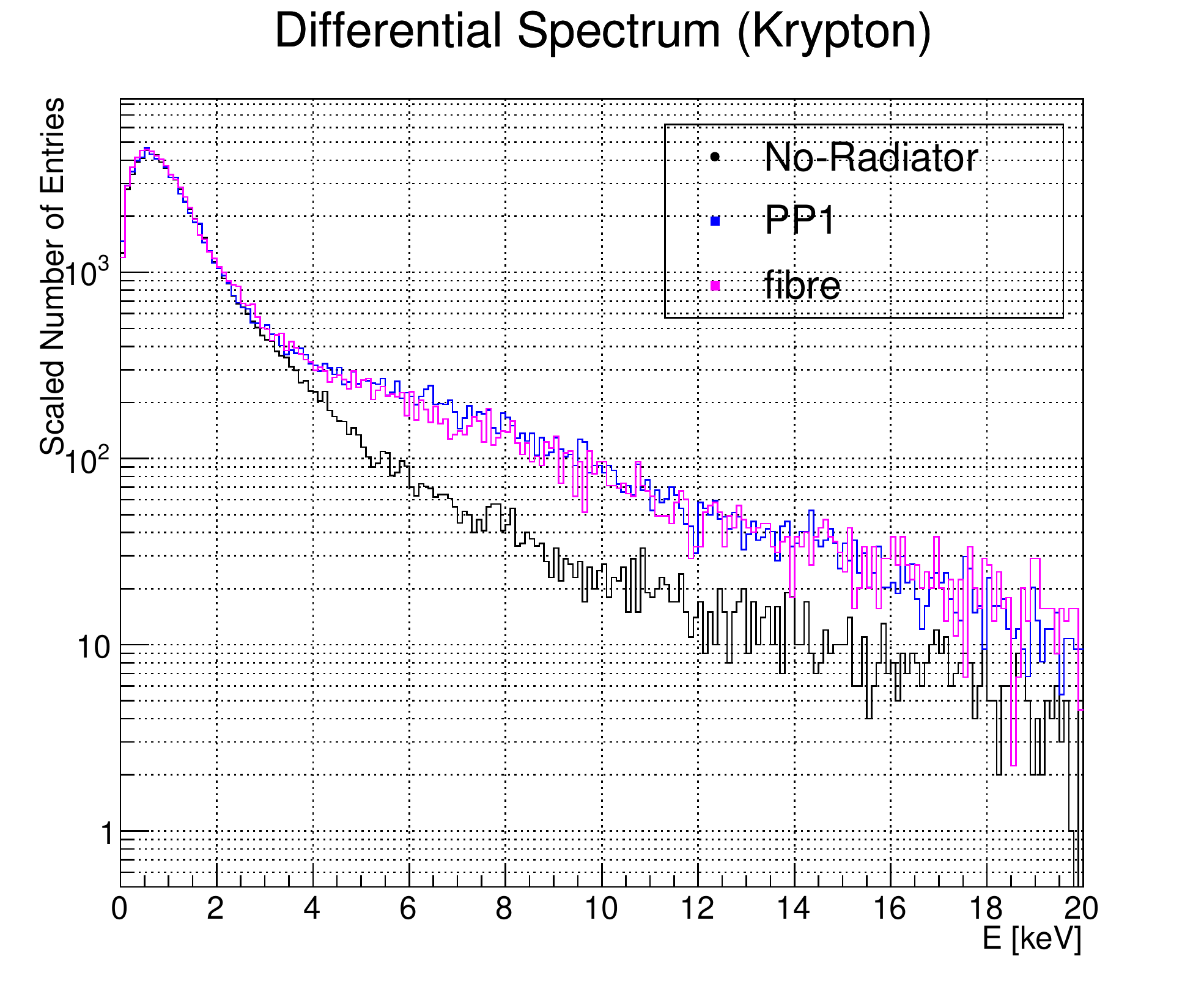}
\caption{\label{diff_kr}Krypton differential spectrum. }
\end{minipage}\hspace{2pc}%
\begin{minipage}{15pc}
\includegraphics[width=15pc]{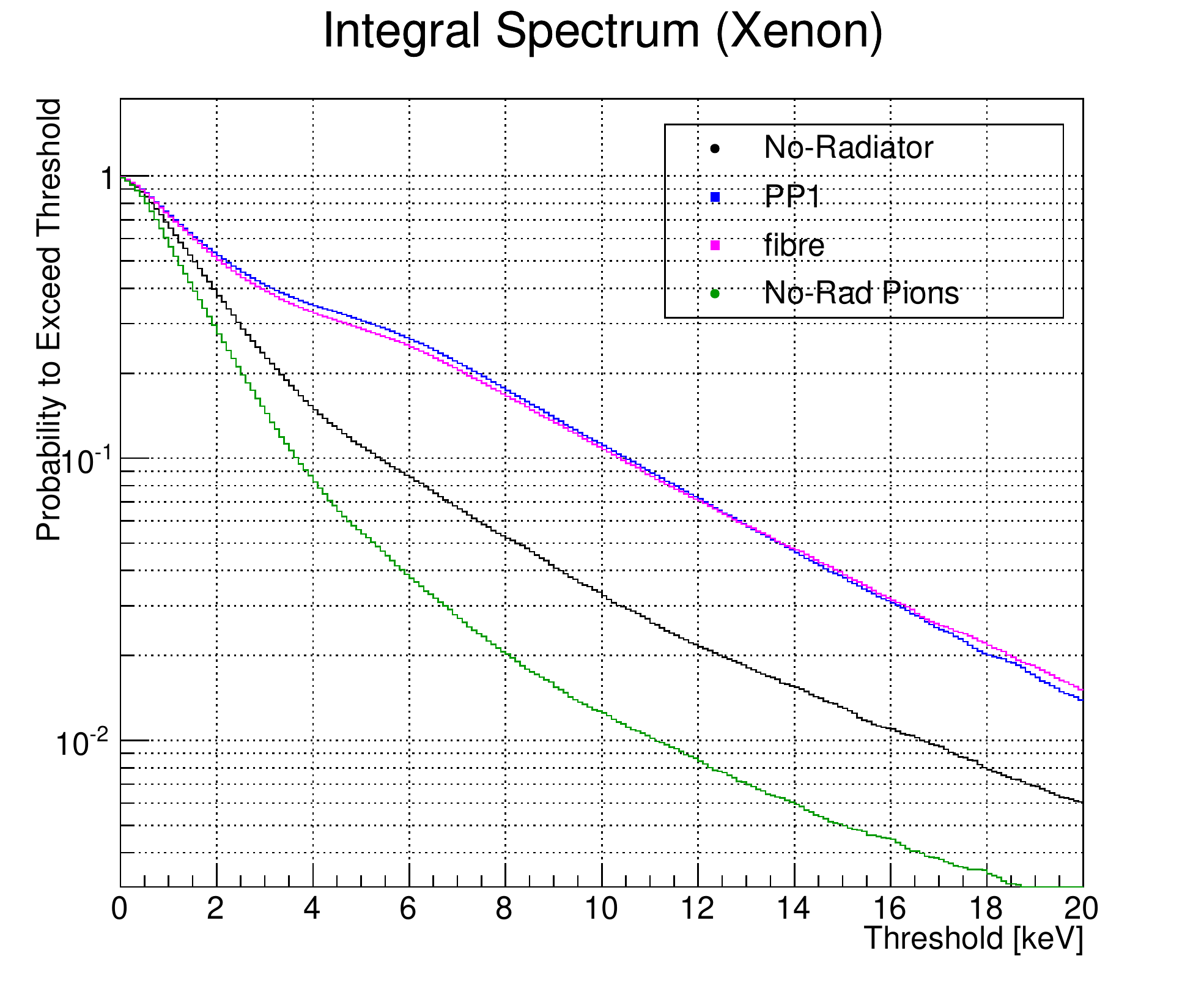}
\caption{\label{int_xe}Xenon integral spectrum. }
\end{minipage}
\end{figure}

 For the differential spectrum histograms we applied a scaling factor to match the low energy parts hence to make an easier comparision(Figures \ref{diff_xe},\ref{diff_ar},\ref{diff_kr}). 
 filled with energy deposited to straw and to make an easy comparison we changed the scaling factor so that the low energy parts of the histograms coincide. By looking at the resulting differential spectrum comparison plots we can infer where TR spectrum starts. A direct observation from these histograms is that the Krypton spectrum slowly varies where Xenon and Argon has small bumps. This might be the result of the remaining energy left after a 12 keV escape photon emmited. 

To investigate various HL thereshold values, we produced integral spectrum plots(Xenon example Figure \ref{int_xe}) by dividing integral result for region after the threshold to total integral for each threshold value. These plots give rejection powers for a specific threshold value. For example, if we set a 6 keV threshold for a single straw filled with Xenon,it would produce HL hit 26\% of the time for 20 GeV electrons however 4\% of the time for pions. 

\section{Results}

  The probability to exceed threshold for electron vs pion plots were produced for different gas and radiator combinations. These plots clearly verifiy that Krypton performs better than Argon but worse than Xenon in all interesting pion rejection powers. 
 
\begin{figure}[h]
\begin{minipage}{15pc}
\includegraphics[width=15pc]{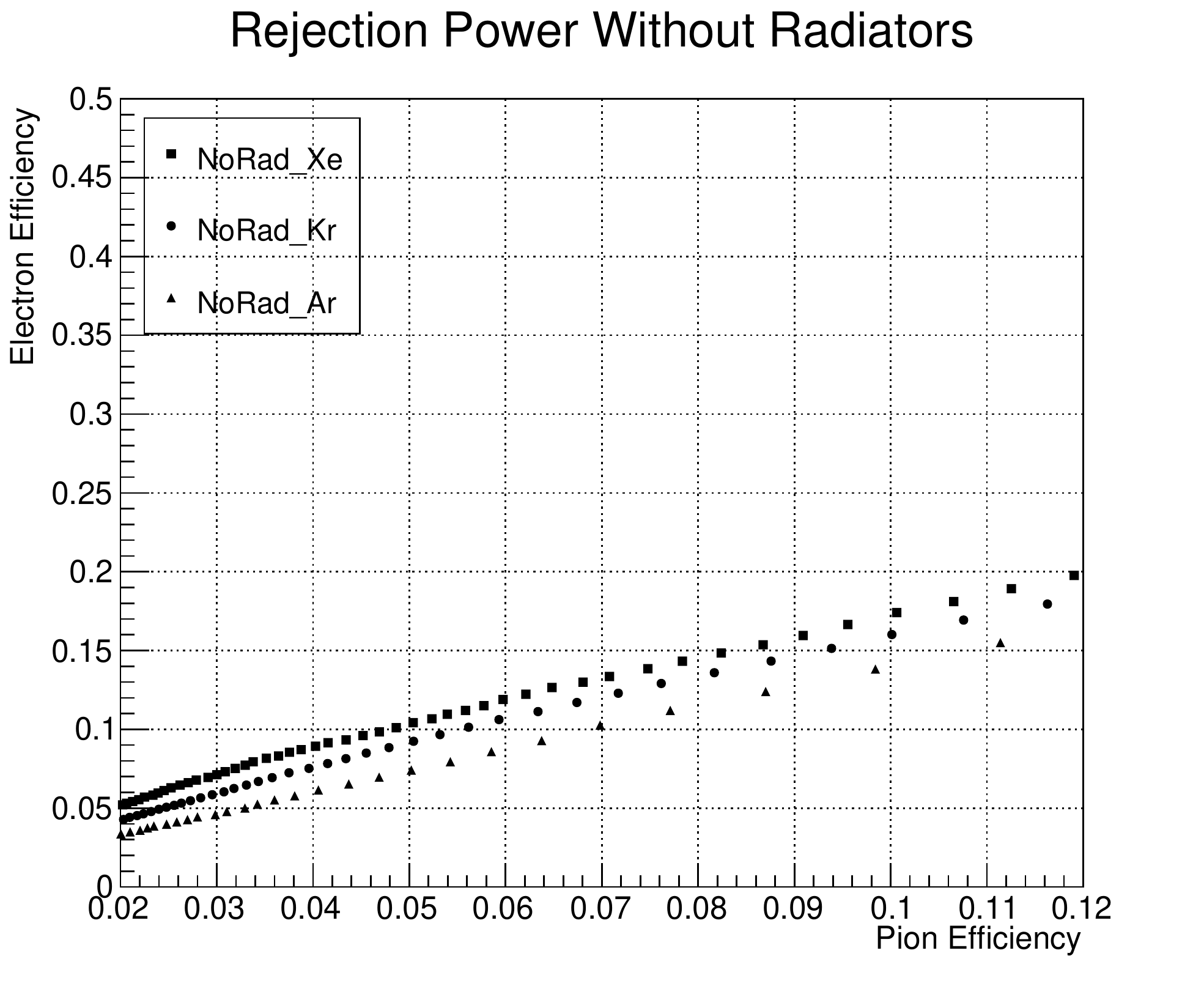}
\caption{\label{fig:norad}Rejection power without radiators for Xe, Ar and Kr. }
\end{minipage}\hspace{2pc}%
\begin{minipage}{15pc}
\includegraphics[width=15pc]{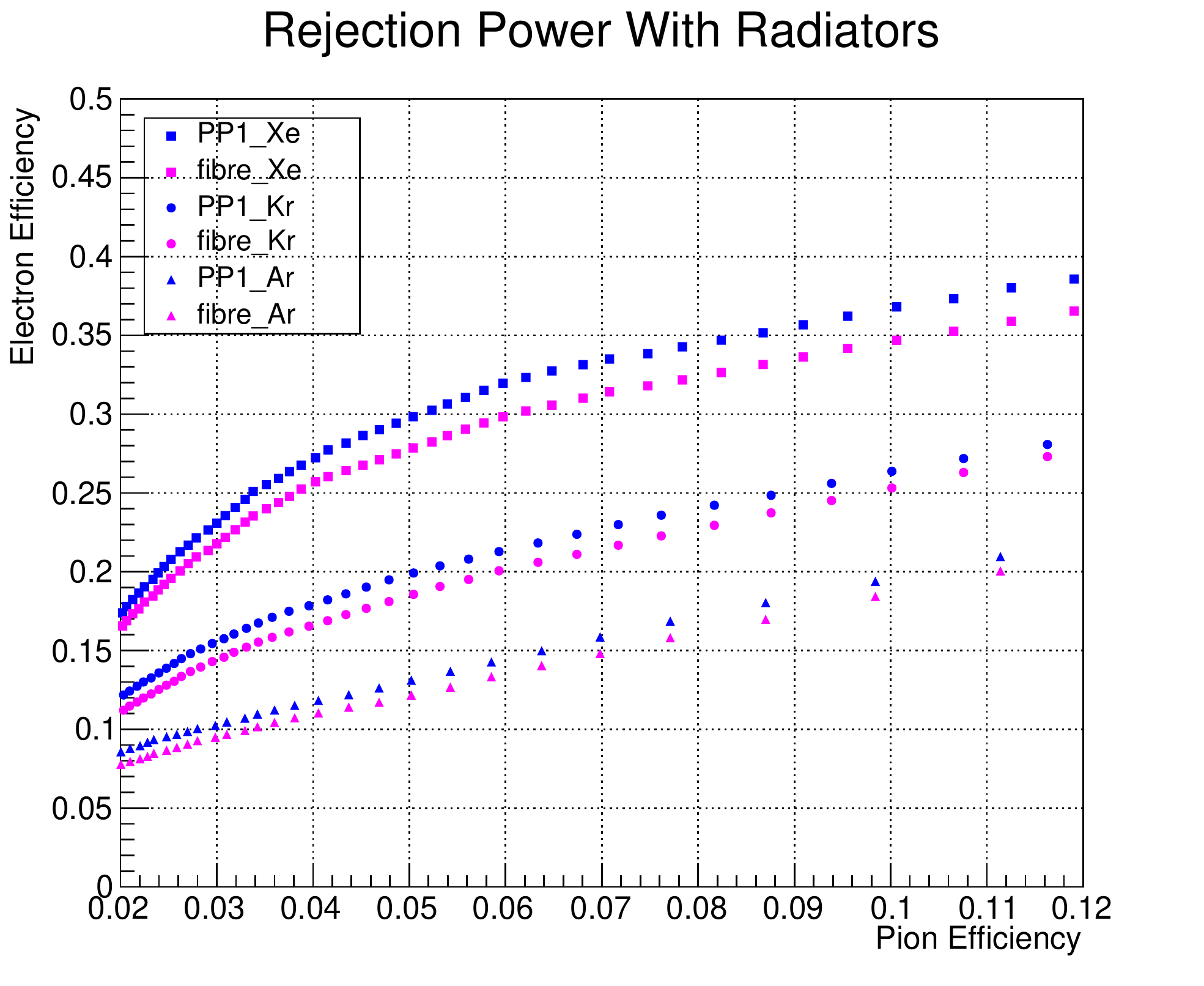}
\caption{\label{fig:rad}Rejection power with radiators for Xe, Ar and Kr. }
\end{minipage}
\end{figure}

 Figure \ref{fig:rad} we can easily make a good estimate of the particle separation of the detectors with the setup used in the test beam. 
 These results are for single straw and using these results one can estimate the pion and electron seperation power for the case of multiple straw layers. 

 The estimated number of straw hits on the particle trajectory can be used to produce binomial distribution along with the probability to exceed threshold for electron and pion values denoted in the Figure \ref{fig:rad}. The resulting distribution will give an seperation estimation. 

\section{Conclusion}

  The analysis performed to obtain HL threshold probability plots will allow us to determine the operating point for TRT straws with alternate gas mixtures. The actual decision for the choice of such operating points require a detailed Monte Carlo simulation of the TRD. 

Although the effect of larger Kr cross section for photon energies greater than 14 keV is not strongly pronounced it's effect on the particle separation it is still quite good. Hence, optimization of the detector geometry for Kr mixture could significantly improve its PID performance. 

\section{Acknowledgments}
We gratefully acknowledge the financial support from Russian Science Foundation – grant No.16-12-10277. We also acknowledge the support from Turkish Atomic Energy Authority(TAEK).


\section*{References}
\begin{thebibliography}{5}
	\bibitem{PDG-2014}
	Olive K A {\it et al. } 2014 {\it Chin. Phys. } C {\bf 38} 090001
	\bibitem{vlad} 
        V O Tikhomirov {\it et al. } Some results of test beam studies of Transition Radiation Detector prototypes at CERN.{\it J. Phys.: Conf. Series} (In this Proceedings)
	\bibitem{graphen}
        A Tishchenko {\it et al. }Effect of graphen monolayer on the transition radiation yield of the radiators based on polyethylene foils. This Conference, http://indico.cfr.mephi.ru/event/4/session/18/contribution/317
        \bibitem{ATLAS-CONF-2011-128}
         ATLAS Collaboration, Particle Identification Performance of the ATLAS Transition Radiation Tracker ,{\it Tech. Rep. ATLAS-CONF-2011-128, CERN, Geneva}
        \bibitem{CXRO}
         The Center for X-Ray optics:X-Ray Attenuation Length, http://henke.lbl.gov/optical\_constants/atten2.html, accessed: 2016-07-01.
\end{thebibliography}
\end{document}